\documentclass[letterpaper, 10 pt, conference]{ieeeconf}  

\IEEEoverridecommandlockouts                              

\usepackage[normalem]{ulem}
\usepackage{amssymb}

\providecommand{\remarkname}{\bf{Remark}}
\usepackage{graphicx}
\usepackage{tikz}

\usepackage{algorithm,algorithmic}
\usepackage{tabularx}
\usepackage[skip=2pt,font=small]{caption}
\usepackage{color,soul}
\usepackage{caption}
\usepackage[position=top]{subfig}
\usepackage{subfig,float}
\usepackage{epstopdf}
\usepackage{diagbox}
\usepackage{multicol}
\usepackage{mathtools}
\usepackage{scalerel}
\usepackage{array} 
\usepackage{booktabs}

\title{\LARGE \bf
Image-Guided Depth Upsampling via Hessian and TV Priors
}

\author{Alireza Ahrabian, Jo$\tilde{\text{a}}$o F. C. Mota and Andrew M. Wallace
\thanks{The authors  are with the School of Engineering and Physical Sciences at Heriot-Watt University (email:
\{aa404, j.mota, a.m.wallace\}@hw.ac.uk).}%
}

\begin{document}

\maketitle
\thispagestyle{empty}
\pagestyle{empty}

\begin{abstract}
We propose a method that combines sparse, depth (LiDAR) measurements with an intensity image to produce a dense, high-resolution depth image. As there are few, but accurate, depth measurements from the scene, our method infers the remaining depth values by incorporating information from the intensity image, namely the magnitudes and directions of the identified edges, and by assuming that the scene is composed mostly of flat surfaces. Such inference is achieved by solving a convex optimisation problem with properly weighted regularisers that are based on the $\ell_1$-norm (specifically, on total variation). We solve the resulting problem with a computationally efficient ADMM-based algorithm. Using the SYNTHIA and KITTI datasets, our experiments show that the proposed method achieves a depth reconstruction performance comparable to or better than other model-based methods.

\end{abstract}

\section{INTRODUCTION}
A key requirement for mobile autonomous systems is the reliable and high resolution estimation of the distances (depth) to surfaces and objects present within a scene. The use of such high resolution ``depth maps" allows enhanced situational awareness that is critical for robotic tasks such as path planning and obstacle avoidance. This is traditionally achieved through the development of computer vision methods that calculate  depth via the joint processing of 3D laser range sensors and RGB camera data. 
\\
\indent Currently, the primary sensors used to obtain reliable depth measurements are scanning laser systems such as the  Velodyne HDL-64e \cite{Geiger2013}. Such sensors tend to provide sparse 3D point cloud data that is very challenging to process due to the unstructured  nature of  the point clouds. As a result,   3D range measurements are generally projected onto a 2D image plane with each pixel of the 2D image corresponding to a depth measurement \cite{Uhrig17}. Such a representation allows us
to combine the dense RGB (intensity) data with the point cloud to recover a high resolution depth map. 
\\
\indent We adopt an optimisation framework as it enables directly encoding assumptions about the problem via regularisers. In particular, we formulate an unconstrained minimisation problem whose objective function has three terms: a data fidelity term that enforces the entries of recovered depth  to conform to the depth measurements, a term that encodes the assumption that the scene is composed of mostly flat surfaces, and another term that incorporates information extracted from the intensity image, mostly information pertaining discontinuities and edges. All these terms are properly weighted according to a heuristic that we devise based on the information given by the intensity image. Thus, the intensity image is used in two different ways: to help identify depth discontinuities at the pixel level, and to weigh the different regularisers. We solve the resulting optimisation problem with an ADMM-based algorithm. Our experiments on the SYNTHIA and KITTI datasets show that the proposed method can reconstruct depth images (with the aid of an intensity image) with an accuracy comparable or better than other model-based approaches

\section{Related Work}
We briefly overview existing methods for image-guided depth upsampling. For a more comprehensive account, see  \cite{Uhrig17}.  Algorithms for  image guided depth upsampling   can generally be divided into  1) local methods based on interpolation filtering, 2) global methods based on global energy function minimisation and 3) learning methods based on convolutional neural networks (CNN's).
\\
\indent \textbf{Local methods}. Apply an appropriately designed  filtering kernel  to the sparse depth measurements \cite{Barron16}\cite{He13}. In particular, the guided image filtering \cite{He13} and joint bilateral filtering \cite{Barron16}  methods utilise the intensity image  to design filtering kernels that result in upsampled depth maps with improved edge preservation. Using more complex structures from the intensity image can further improve the performance of local methods. For example, the work in \cite{Liu2013} uses geodesics to vary the support of the filter kernel, which improves the fidelity of the upsampled depth image.    
\\
\indent \textbf{Global methods}. In contrast to local methods, global methods consist of algorithms that solve optimisation problems whose terms encode prior knowledge about the scene. Such prior knowledge typically takes the form of a global measure, for example, that the image is smooth or has a small number of edges. For example, the work in \cite{Ferstl13} proposed a method with regularization based on total generalised variation (TGV), which promotes sharp edges by using a weighting scheme based on an anisotropic diffusion tensor.  Other methods incorporate more sophisticated assumptions. For example, the work in \cite{Schneider16} includes semantics learned from the intensity image, while  \cite{Song19}  learns a dictionary that couples the ideal depth map and the intensity image. This learned dictionary is then utilised in an optimisation problem to recover the depth image. Finally, a nonconvex formulation has also been proposed \cite{Ham18} that incorporates the guiding intensity image via a nonconvex regulariser. 
\\
\indent \textbf{Learning methods}. Have been proposed   for image guided depth upsampling  \cite{Uhrig17}. For example, Hui et al. \cite{Hui16} used three sub-networks to  extract features from both the depth and intensity image. These are then fed into a fusion stage to jointly process the features and thus output an upsampled depth image. 
\\
\indent \textbf{Discussion}. Local methods have drawbacks that include inconsistent local correlations between the intensity image and sparse depth measurements. This may result in an inadequate upsampled depth map. Global methods overcome limitations that arise with local  methods by minimising an energy functional over the entire depth map, but a difficulty still arises in defining an effective regulariser (that incorporates the guiding intensity image). In particular, the weighted TGV regulariser in \cite{Ferstl13} only informs the optimisation of a change in depth via the anistropic diffusion tensor. Such a formulation does not include information regarding the magnitude of the depth change, implying that the method  can be effected by strong intensity textures. The work in \cite{Ham18} overcomes this limitation by introducing a more sophisticated regulariser that encodes relevant  structural relations (between the intensity image and depth map). However, this results in a non-convex optimisation problem that can only guarantee locally optimal solutions.
While learning methods  have shown superior performance \cite{Uhrig17}  over traditional  methods (i.e. local and global techniques),  this is achieved through the use of highly parameterised statistical models that can potentially fail if the training data does not capture all possible scenarios that may arise. 
\\
\indent Hence, in this work we propose a weighted regulariser incspired by the work in \cite{Mota17}. In particular, our formulation enables us to include the magnitude of  the change in depth (obtained from the intensity image) whilst guaranteeing global convergence.   
\begin{figure}[!t]
\centering
\includegraphics[trim=0 100 0 100,clip,width=1\columnwidth]{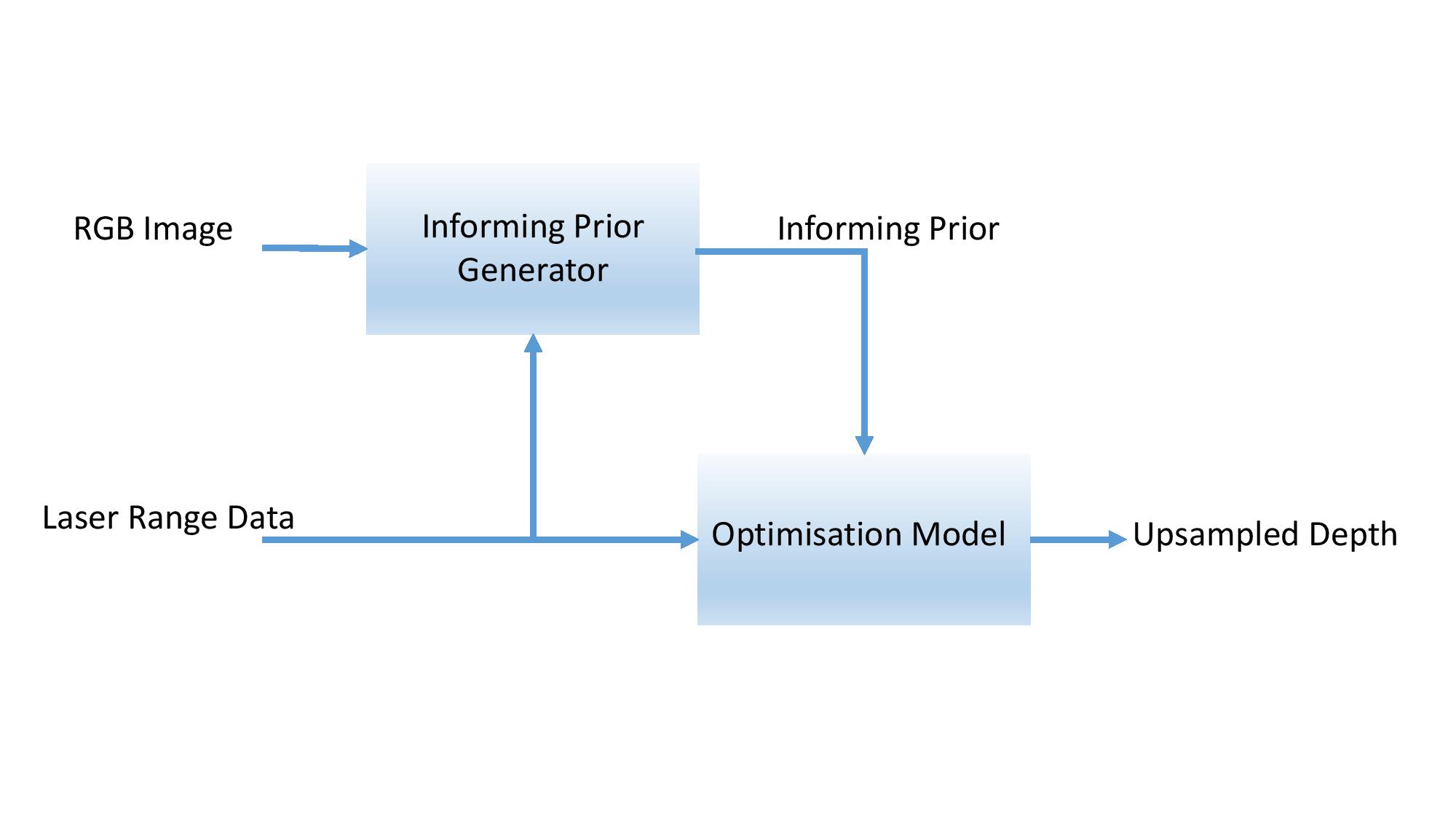}
\caption{Proposed workflow for depth reconstruction using both the RGB image and laser range data. The range data and the intensity image generate an informing prior, where this is used to recover an upsampled depth image.
 }\label{fig:Model}
\end{figure}
\section{Proposed Method}
In this work we generate a high resolution depth map $X\in \mathbb{R}^{M\times N}$ (where $M$ and $N$ respectively correspond to the number of rows and columns in the image), by using both 3D data points obtained from a laser range sensor $L \in \mathbb{R}^{3\times M_{L}}$ (where $M_{L}$ is the number of data points) and an intensity image $I_{C}\in \mathbb{R}^{M\times N}$ obtained from an RGB camera. The laser range data points are projected onto the camera reference frame, resulting in the input depth measurements  $L_{C}\in \mathbb{R}^{M\times N}$. Finally, $L_{C}$ is a $\mathit{k}$-sparse matrix with $\mathit{k}\ll MN$.
\\
\indent Our method was inspired by the theory of $\ell_1$-$\ell_1$ minimisation \cite{Mota17}\cite{Mota2017}, which provides guarantees for sparse reconstruction in the presence of side information (an informing prior), i.e., a signal similar to the signal to reconstruct. The theory, however, does not apply to our specific formulation, detailed next, as we had to consider more complex priors.  To this end, our depth upsampling problem  (Fig.~\ref{fig:Model}) can be split into the following sub-problems: 1) form an optimisation problem that utilises a weighted regulariser based on the $\ell_{1}-\ell_{1}$ penalty terms. The input data includes both the  transformed laser range data $L_{C}$ as the under-sampled depth measurements along with the informing prior $U$, determined from  intensity image and coarse estimate of the sparse depth measurements. 2) Propose a method for generating the informing prior $U$ via a heuristic transformation that   utilises both  the image intensity data $I_{C}$ and a coarse estimation of the depth image $L_{C}$. The proposed transformation enables us to construct an informing prior that is more robust to changes in intensity caused by strong  textures and shadow effects. 
\begin{figure}[!t]
\captionsetup[subfigure]{}
\centering
\subfloat[]{\includegraphics[width=0.5\columnwidth]{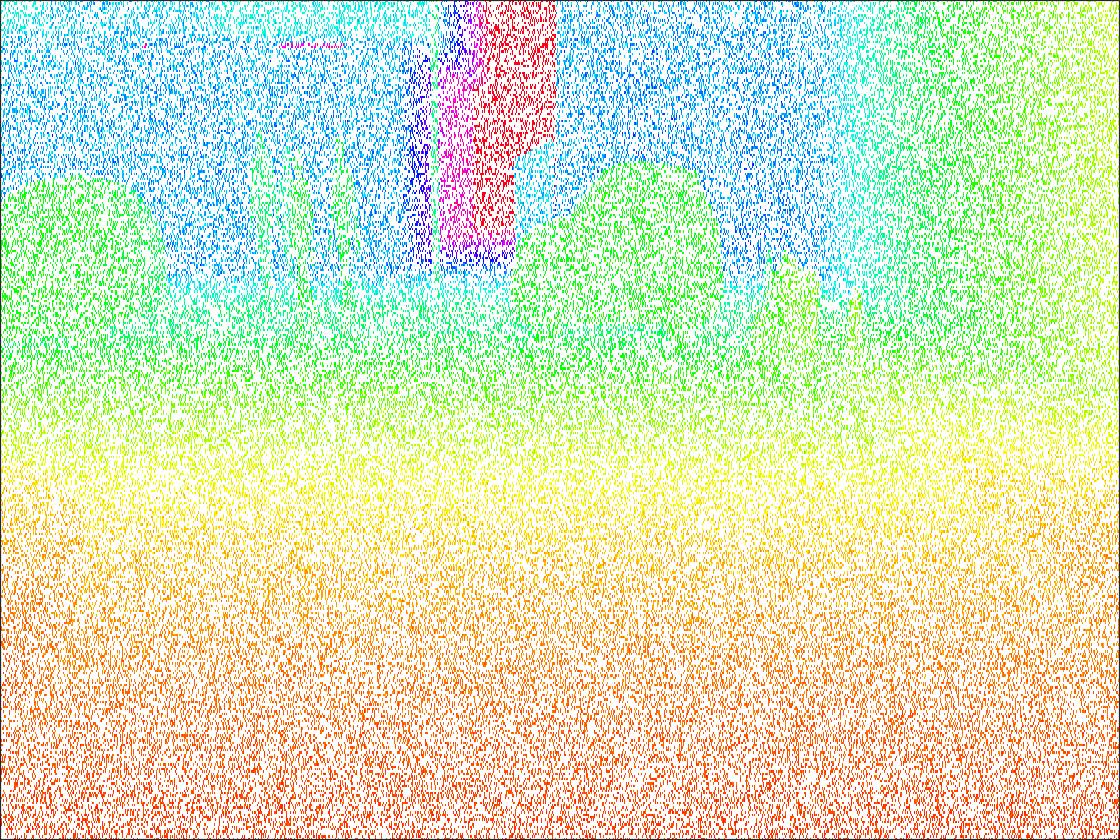}\label{fig:rawdata1}}
\subfloat[]{\includegraphics[width=0.5\columnwidth]{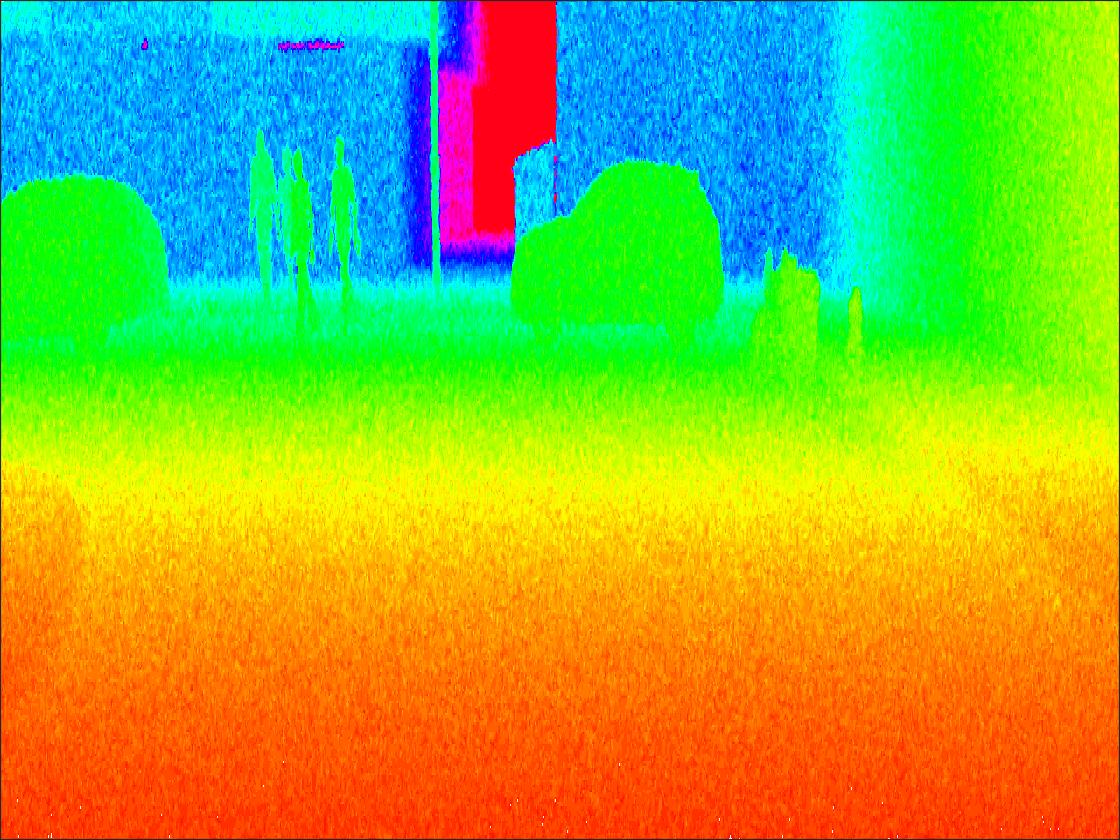}\label{fig:rawdata2}}
\\
\vspace{-3mm}
\subfloat[]{\includegraphics[width=0.5\columnwidth]{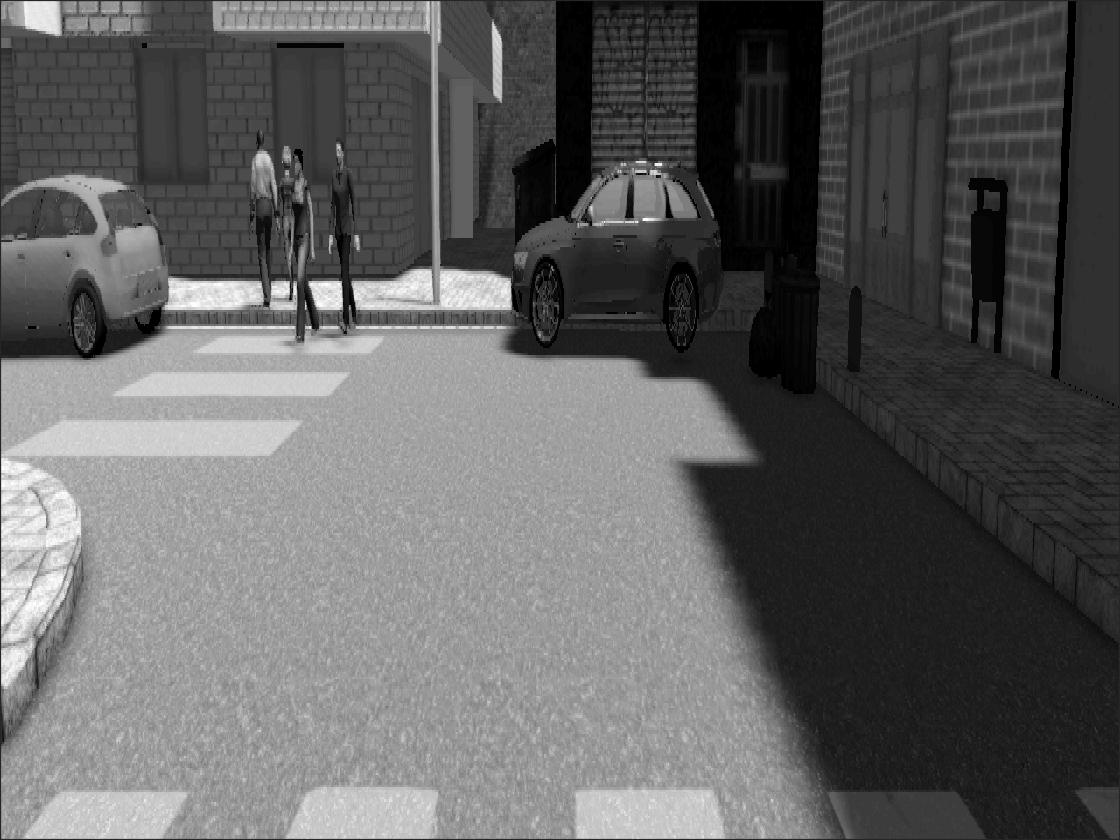}\label{fig:rawdata3}}
\subfloat[]{\includegraphics[width=0.5\columnwidth]{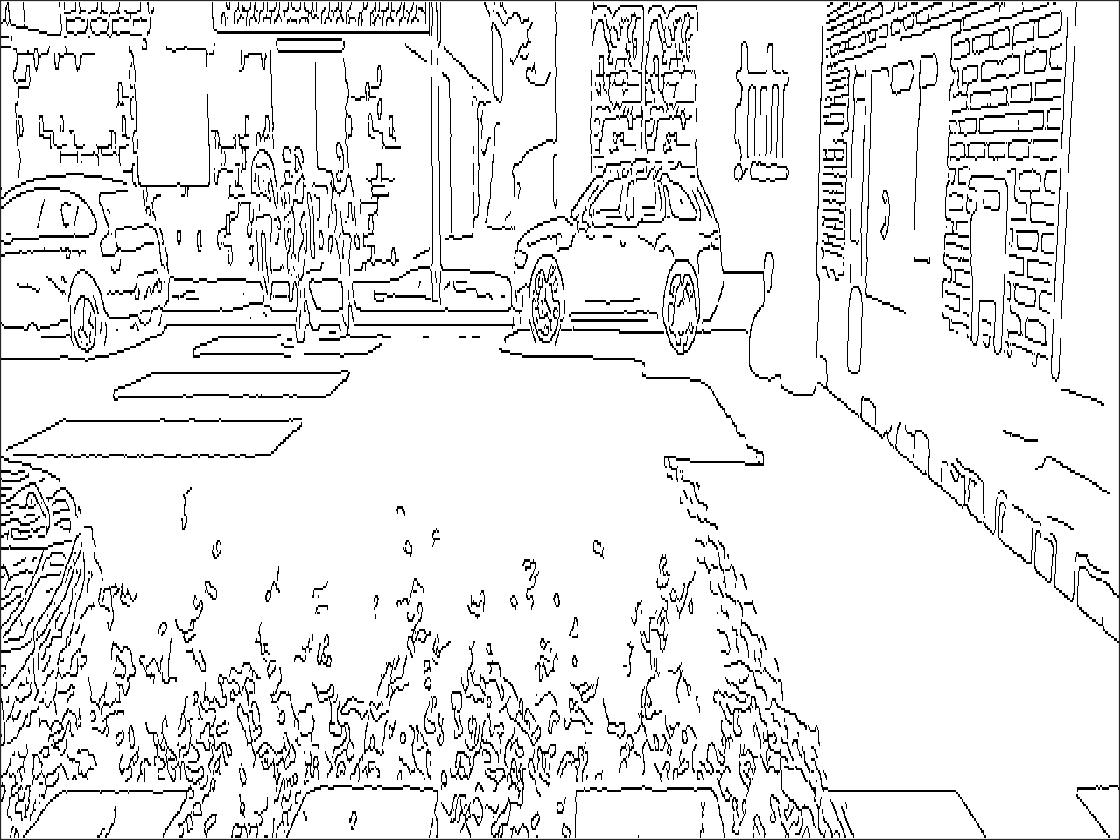}\label{fig:rawdata4}}
\caption{Information used for generating the informing prior (SYNTHIA dataset). (a)  Camera projected laser range data, (b) a coarse upsampled depth map, (c)  the intensity image  and (d) the  extracted intensity discontinuities (edges).   }
\label{fig:rawdata}
\end{figure}
\subsection{Depth Recovery Model}
Our optimisation strategy for recovering  the upsampled depth image $X$ uses both the sparse camera-projected laser range measurements $L_{C}$ and the dense camera intensity data $I_{C}$. Let $x\in \mathbb{R}^{n}$ and $b\in \mathbb{R}^{n}$ (where $n=MN$) correspond to the  column-wise vectorisation of  $X$ and  $L_{C}$ respectively. The proposed method for determining $x$ is given by the solution to the  following optimisation problem  
\begin{equation}
 \begin{aligned}
&\underset{x}{\text{minimise}}\quad\hspace{-1.2mm}\frac{1}{2}||Sx-b||_{2}^{2}+\mathcal{\hat{R}}(x,u;W_{1},W_{2})
\end{aligned}
\label{OptProb}
\end{equation}
where $\mathcal{\hat{R}}()$ is the weighted regulariser,  $S\in \mathbb{R}^{n\times n}$ is a diagonal matrix such that $s_{i,i}=1$ if a data sample has been acquired and  $s_{i,i}=0$ otherwise, and $u\in \mathbb{R}^{n}$ corresponds to the vectorised  side information $U$ that is generated from the intensity data $I_{C}$ (a detailed explanation is provided in Section IV.C). The diagonal matrix $W_{j}\in \mathbb{R}^{2n \times 2n}$ for $j\in\{1,2\}$, with each element $0\leq w^{j}_{i,i}\leq 1$,  encodes the confidence in the  support (at index $i$) used for reconstructing the depth  as in \cite{Weizman2014CompressedSF}. The term $\mathcal{\hat{R}}(x,u;W_{1},W_{2})$ is given by
\begin{equation}
\mathcal{\hat{R}}(x,u;W_{1},W_{2})=\beta||W_{1}Hx||_{\text{1}}+\gamma||W_{2}D(x-u)||_{1}  
\label{Reuglariser}
\end{equation}
where the first term  enforces slowly varying changes in depth, and the second term  penalises differences between the reconstructed depth image and the information $u$ extracted from the intensity image. We define the sparsity induced by the first term as the   Hessian total variation (HTV) (please refer to example 2.2 in  \cite{Setzer11} for a definition), and is determined by a linear transformation of the vectorised depth using the circulant matrix $H=[H^{T}_{r},H^{T}_{c}]^{T}\in  \mathbb{R}^{2n\times n}$; where $H_{r}\in   \mathbb{R}^{n\times n}$ is the second difference  along the rows while $H_{c}\in  \mathbb{R}^{n\times n}$ is the second difference along the columns. The sparsity induced by the second term is defined as anisotropic total variation (TV) \cite{Liu15}, and is determined by a linear transformation of the vectorised depth using the circulant matrix   $D=[D^{T}_{r},D^{T}_{c}]^{T}\in  \mathbb{R}^{2n\times n}$; where $D_{r}\in   \mathbb{R}^{n\times n}$ is the first difference  along the rows while $D_{c}\in  \mathbb{R}^{n\times n}$ is the first difference along the columns. 
\\
\indent A particularly advantageous feature of the proposed  approach is that the informing prior $Du\in  \mathbb{R}^{2n}$ (with $u=g(I_{C})$, where $g(\cdot)$ is a  functional mapping that needs to be determined)  does not need to exist in the same domain as the measurements, e.g. when compared to the weighting approach in \cite{Ferstl13} incorporating the intensity image for depth upsampling guidance. Our approach based on an informing prior enables more information (magnitude and direction of changes in depth) to be included in the optimisation process, potentially provides more accurate reconstruction of the depth map. Finally, the constants $\beta$ and $\gamma$ are the model hyperparameters.  
\subsection{ADMM Solver}
We use the alternating direction method of multipliers (ADMM) to efficiently solve the proposed optimisation problem  \eqref{OptProb}. In particular, we follow the parametrisation strategy outlined in \cite{Liu15}. That is,  auxiliary variables are introduced such that   matrices  with similar properties (i.e. circulant or diagonal) can be grouped together to solve the proximal operators. Following \cite{Liu15}, we recast \eqref{OptProb} as
\begin{equation}
 \begin{aligned}
&\underset{x,r,l,p,v,g,z,k}{\text{minimise}}\quad\hspace{1.2mm}\frac{1}{2}||Sx-b||_{2}^{2}+\beta||p||_{1}+\gamma||k||_{1}  \\
&\text{subject to}\quad x=r,\hspace{1mm} v=Hr,\hspace{1mm} l=v,\hspace{1mm} p=W_{1}l\\ 
&\hspace{1.7cm}g=Dr-Du,\hspace{1mm} z=g,\hspace{1mm} k=W_{2}z
\end{aligned}
\label{OptProbParam}
\end{equation}
where $\{x=r, v=Hr, l=v, p=W_{1}l, g=Dr-Du, z=g, W_{2}z=k \}$ correspond to the set of equality constraints that need to be satisfied, given the  auxiliary variables $r,l,p,v,g,z,k$. The problem in \eqref{OptProbParam} can be solved efficiently, as solutions to  proximal operators require the inversion of either diagonal   or circulant matrices; where solutions for proximal operators with circulant matrices can be efficiently calculated using the fast Fourier transform. Please refer to \cite{Liu15}  for exact implementation details, given the parameterised model \eqref{OptProbParam}.
\subsection{Informing Prior Generation}
So far we have presented an optimisation model that uses   both  sparsity in depth  and sparsity with respect to an  informative prior.  However, we have yet to present a principled approach  for generating  such a prior. To guide our approach, we review a result \cite{Mota17} regarding the `quality' of such a prior. There, it was shown  that provided we have a sufficient number of  overlapping support points between the side information and the ideal signal,   then given a fixed number of sensor measurements, incorporating the side information improves the reconstruction error (in the mean squared error sense). A support point is defined as follows, $\{i :|x_{i}|>0, \quad x_{i}\in x, \quad x\in \mathbb{R}^{n} \}$.
\\
\indent To this end, our approach seeks to design an informative prior, $\hat{u}=Dg(I_{C})$, such that the number of overlapping support points  with respect to first difference of the ideal vectorised depth map $Dx^{\star}$  is maximised. Given  that the $\ell_{1}$-norm of first differenced signals promotes sparsity with respect to discontinuous changes that arise in the signal; we can observe the following relationship between the intensity image and depth map. Namely, a significant  proportion of depth discontinuities  coincide with  discontinuities in  intensity. Informed by this relationship between  $X$ and $I_{C}$, we generate $\hat{u}$  as follows: 
\\
\textbf{Step 1} - Define $\hat{U}_{r}\in\mathbb{R}^{M\times N}$ and $\hat{U}_{c}\in\mathbb{R}^{M\times N}$, the respective row and column wise first differences of the prior $U$. Furthermore,  upsample the camera-projected laser range measurements $L_{C}$ (see Fig.~\ref{fig:rawdata1} for example), to obtain a (relatively) coarse estimate of the depth map $X_{i}$ (see Fig.~\ref{fig:rawdata2}). Examples of interpolation methods used in this work include the methods outlined in \cite{Hawe11} and \cite{Ferstl13}. 
\\
\textbf{Step 2} - Identify discontinuities in the intensity image $I_{C}$ (see Fig.~\ref{fig:rawdata3}). In this work we have used the Canny edge detector. The pixel co-ordinates of the discontinuities are stored in $E_{C}\in\mathbb{R}^{M\times N}$, that is a  matrix with binary-valued  entries (see Fig.~\ref{fig:rawdata4}). 
\\
\textbf{Step 3} - For each pixel co-ordinate $(r_{co},c_{co})$ in $\hat{U}_{r}$, $\hat{U}_{c}$, $E_{C}$ and $X_{i}$,  determine if it is an edge co-ordinate using $E_{C}$. 
\\
\textbf{Step 4} - If an edge is present at  $(r_{co},c_{co})$,   estimate the depth  adjacent (along the rows and columns) to the edge as follows
\begin{equation}
 \begin{aligned}
&d_{l}=\text{median}(x^{i}_{(r_{co}-M_{p},c_{co})},...,x^{i}_{(r_{co}-1,c_{co})}) \\
&d_{r}=\text{median}(x^{i}_{(r_{co}+1,c_{co})},...,x^{i}_{(r_{co}+M_{p},c_{co})}) \\
&d_{u}=\text{median}(x^{i}_{(r_{co},c_{co}+1)},...,x^{i}_{(r_{co},c_{co}+M_{p})}) \\
&d_{d}=\text{median}(x^{i}_{(r_{co},c_{co}-M_{p})},...,x^{i}_{(r_{co},c_{co}-1)}) 
\end{aligned}
\label{DepthEst}
\end{equation}
where  $x^{i}_{(\cdot,\cdot)}$ is an element of $X_{i}$, $M_{p}$ the number of pixels adjacent to the edge and $\text{median}(\cdot)$ corresponds to the median estimator. The estimates $d_{l}$ and $d_{r}$ correspond to adjacent depth estimates left and right (respectively) of the edge; while,  $d_{u}$ and $d_{d}$ correspond to the adjacent depth estimates above  and below  (respectively) the edge.
\\
\textbf{Step 5} - Finally, given that  $\hat{u}^{r}_{(\cdot,\cdot)}$ and $\hat{u}^{c}_{(\cdot,\cdot)}$ are the respective  elements of the matrices $\hat{U}_{r}$ and $\hat{U}_{c}$, we determine  the value of the depth discontinuities  for all edge co-ordinates as follows
\begin{equation*}
\hat{u}^{r}_{(r_{co},c_{co})}=
  \begin{cases}
d_{r}-d_{l}     & \quad   |d_{r}-d_{l}|>0\\
 0 & \quad \text{otherwise}
  \end{cases}
  \label{c1}
\end{equation*}
\begin{equation*}
\hat{u}^{c}_{(r_{co},c_{co})}=
  \begin{cases}
d_{d}-d_{u}     & \quad   |d_{d}-d_{u}|>0\\
 0 & \quad \text{otherwise}
  \end{cases}
  \label{c2}
\end{equation*}
 For all pixel co-ordinate that do not correspond to an edge, $\hat{u}^{r}_{(\cdot,\cdot)}=\hat{u}^{c}_{(\cdot,\cdot)}=0$. An estimate of $\hat{u}$ is obtaining by vectorising $\hat{U}=[\hat{U}_{r},\hat{U}_{c}]$.    
\\
\\
\textbf{Remark.} \textit{Extracting the location of the intensity discontinuities from  the  intensity data and estimating the corresponding change in depth results in following advantages: 1) If the intensity discontinuity  location is within the vicinity of a true depth discontinuity, then the estimated change in depth is likely to enhance the reconstruction. 2) If the intensity discontinuity is caused by a texture change with negligible change in the true depth, then the estimated change in depth is likely to be small and therefore less likely to affect the reconstruction. }
\\ 
\begin{table*}[!t]
\centering
\normalsize
\begin{tabular}{@{\extracolsep{10pt}}lcccccc@{}}
\toprule
\multicolumn{1}{c}{}&\multicolumn{2}{c}{\textit{Old European City}}& \multicolumn{2}{c}{\textit{Highway}} & \multicolumn{2}{c}{\textit{New York Like City}}\\
\cline{2-3}
\cline{4-5}
\cline{6-7}
    & 6.25\% & 1.56\%  & 6.25\% & 1.56\% & 6.25\% & 1.56\% \\
\midrule
Proposed Method      & \textbf{0.35}/\textbf{1.77}    & 0.62/\textbf{2.47}     &0.29/1.69     & 0.56/2.57  & \textbf{0.40}/\textbf{1.94}    & \textbf{0.70}/\textbf{2.72}       \\
TGV \cite{Ferstl13}       &0.37/1.80    & 0.73/2.49     & \textbf{0.28}/\textbf{1.57}    & 0.65/\textbf{2.36}     & 0.42/2.06   & 0.73/3.03   \\
SD Filter \cite{Ham18}       & 0.39/1.94     & \textbf{0.60}/2.69     & 0.30/1.68    & \textbf{0.49}/2.53     & 0.45/2.20    & 0.82/2.79   \\
Nearest Neighbour   & 0.67/2.32&   0.87/3.07    &   0.63/2.40  &  0.85/3.38 & 0.72/2.50   & 0.94/3.33  \\
\bottomrule
\end{tabular}
\caption{Evaluation of the proposed method using the SYNTHIA dataset. The results are carried out using three sequences at two different sparse sampling levels: $6.75\%$ and $1.56\%$. The error (in metres) is measured by the MAE and RMSE (MAE/RMSE), along with the best result being highlighted.  }
\label{tab:Table1}
\end{table*}
\vspace{-2mm}
\begin{figure*}[!t]
\captionsetup[subfigure]{}
\centering
\subfloat[Groundtruth]{\includegraphics[width=0.6\columnwidth]{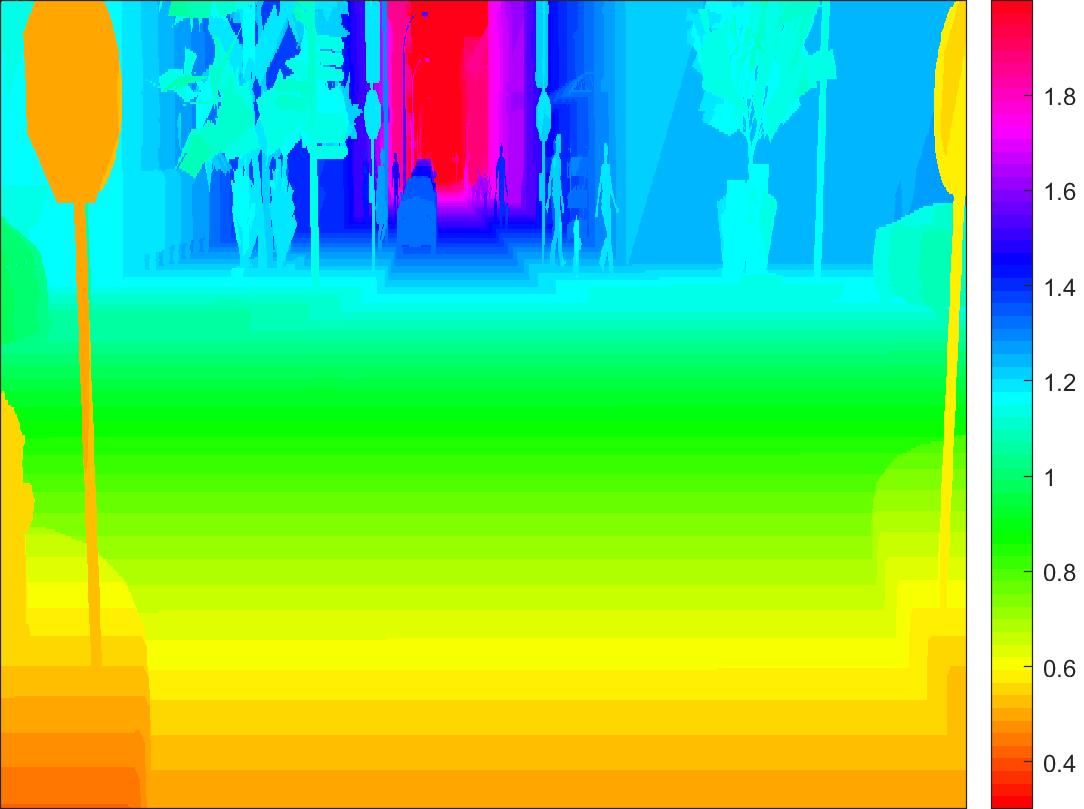}\label{fig:SynthiaRes_GT}}
\quad
\subfloat[RGB Image]{\includegraphics[width=0.6\columnwidth]{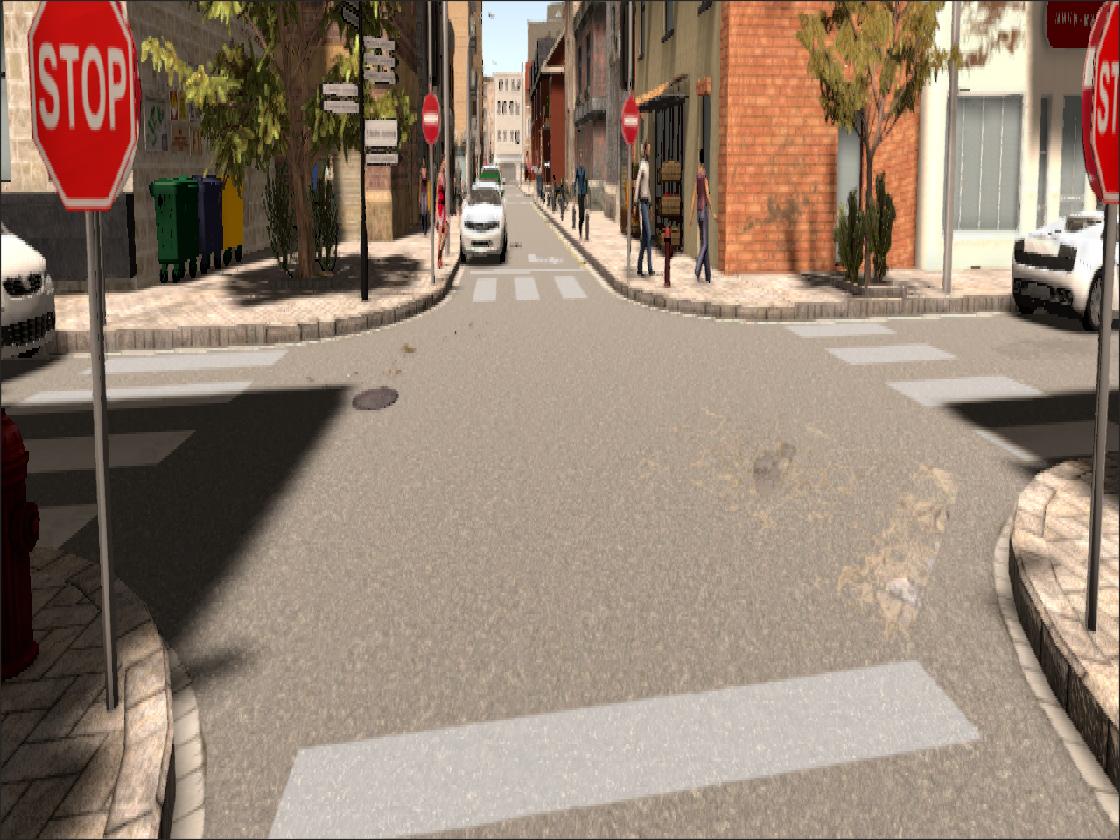}\label{fig:SynthiaRes_Int}}
\quad
\subfloat[TGV \cite{Ferstl13}]{\includegraphics[width=0.6\columnwidth]{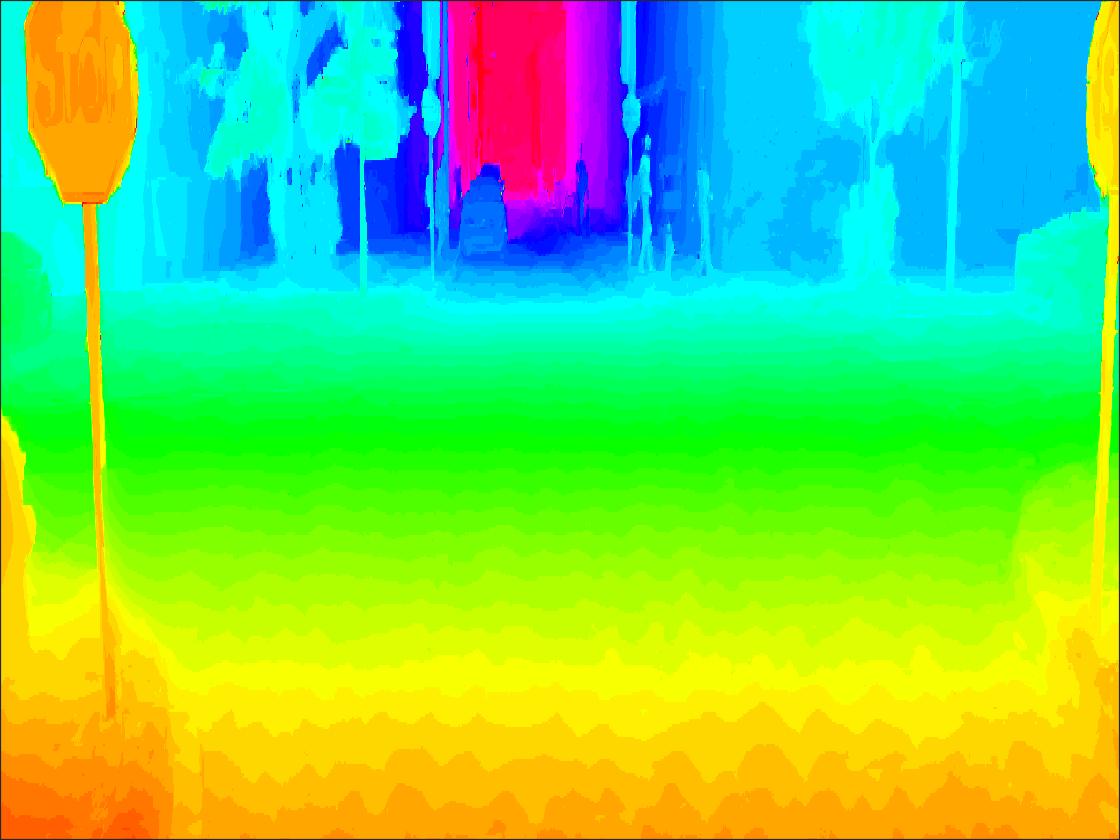}\label{fig:SynthiaRes_TGV}}
\\
\vspace{-2mm}
\subfloat[SD Filter \cite{Ham18} ]{\includegraphics[width=0.6\columnwidth]{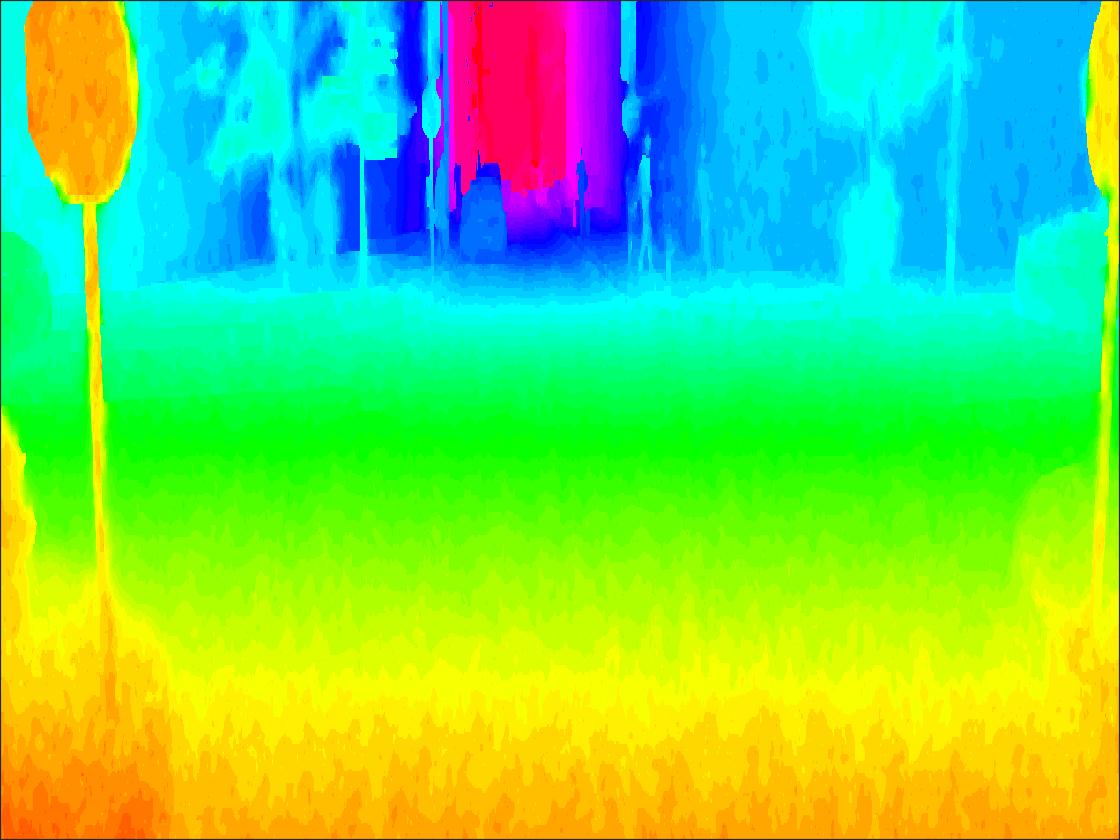}\label{fig:SynthiaRes_SD}}
\quad
\subfloat[Proposed Method]{\includegraphics[width=0.6\columnwidth]{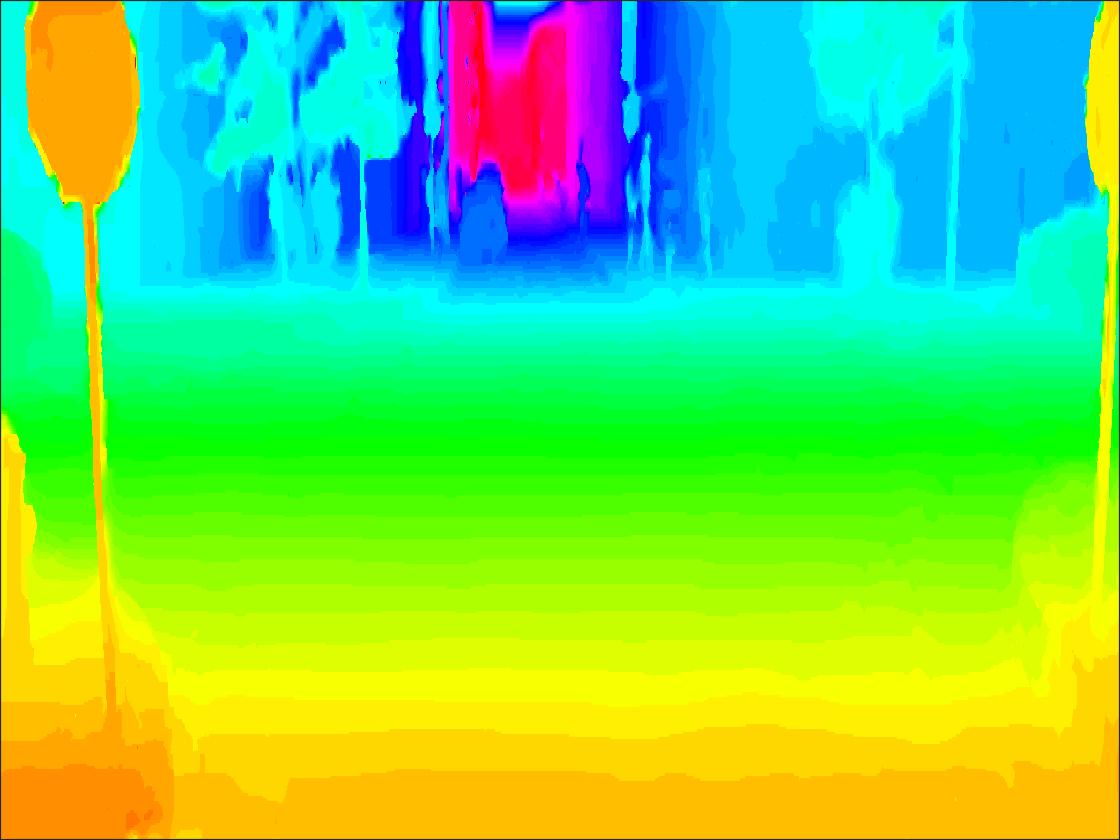}\label{fig:SynthiaRes_PM}}
\quad
\subfloat[Baseline (without informing prior term)]{\includegraphics[width=0.6\columnwidth]{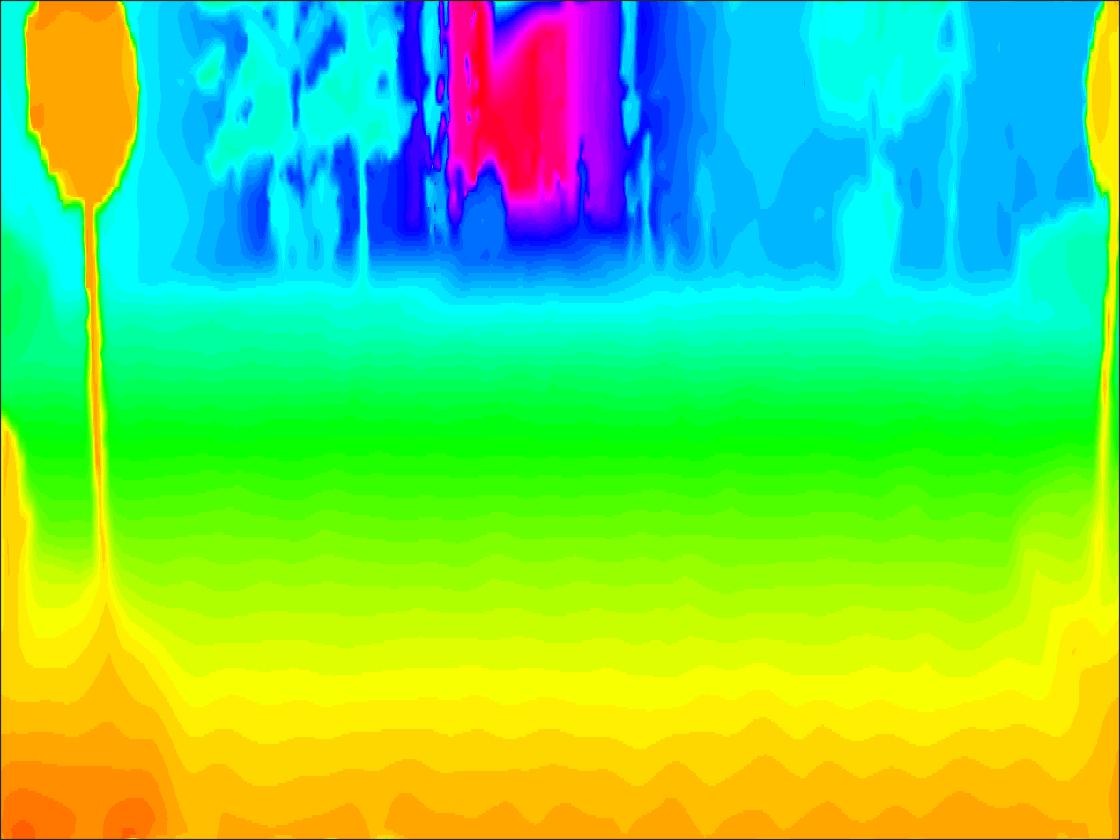}\label{fig:SynthiaRes_Base}}
\caption{A qualitative comparison of a single instance of the  \textit{Old European City} sequence with a measurement  sparsity level of $6.25\%$. (a) Groundtruth depth. (b) RGB image. (c) TGV \cite{Ferstl13}. (d) SD filter \cite{Ham18}. (e) Proposed method.  (f) Baseline depth upsampling (basic optimisation model that only utilises the Hessian TV regularisation term in \eqref{OptProb}). The logarithm of the depth values were used for colouration.}
\label{fig:SynthiaRes}
\end{figure*}
\subsection{Weigh Selection}
The weighting terms $W_{1}$ and $W_{2}$ effectively encode the confidence in the respective sparse penalty terms (that is, either the Hessian TV or TV regularisers) \cite{Weizman2014CompressedSF}. However, to design an effective weighting scheme, we first consider the following scenarios that may arise regarding the proposed regulariser \eqref{Reuglariser}. Namely, for each element of the informing prior $\hat{u}$, we would either have a nonzero value corresponding to information that will guide depth recovery, or a zero value indicating no guidance is available. This leads to the following cases:
\begin{enumerate}
\item \textit{No informing prior is available} - In this case, we do not have any information  that would lead us to enforce a higher demand for sparsity/regularisation from  either the Hessian TV or TV terms.
\item \textit{Informing prior available} - The presence of an informing prior at a given pixel indicates a strong chance of a depth discontinuity at that pixel. As a result, we  enforce a  weight on the TV term smaller (i.e., stronger) than the weight of the Hessian TV term. 
\end{enumerate}
\indent Based on the statements above, we design the following weighting schemes for both $W_{1}$ and $W_{2}$ using the informing prior vector $\hat{u}$. That is, for each diagonal element  $w^{1}_{i,i}$ in $W_{1}$, we set
\begin{equation*}
w^{1}_{i,i}=
  \begin{cases}
0    & \quad   if \quad |[\hat{u}]_{i}|>0 \quad or  \quad |[\hat{u}]_{i+1}|>0\\
 1 & \quad else
  \end{cases}
  \label{c1}
\end{equation*}
\begin{equation*}
w^{1}_{j,j}=
  \begin{cases}
0     & \quad   if \quad |[\hat{u}]_{j+M}|>0 \quad or  \quad |[\hat{u}]_{j+M}|>0\\
 1 & \quad else
  \end{cases}
  \label{c2}
\end{equation*}
for $1<i<n$ and $n<j<2n$, while $W_{2}=I_{2n}$, where $I_{2n}$ is the identity matrix with dimension $2n$. The operator  $[\cdot]_{i}$ corresponds to the $i^{th}$ element of a vector. The selection of the weighting term $W_{1}$ arises from the second difference of a step change that results in two non-zero values (while one non-zero value is produced when calculating the first difference of a step change).
\section{Experimental Results}
\subsection{\text{SYNTHIA} Dataset}
We first evaluate the performance of our proposed method on the SYNTHIA dataset \cite{Ros_2016_CVPR}. The primary advantages of this dataset are as follows, 1) the variation in the sequence scenarios being recorded (i.e. city or highway), 2) the control over sparsity levels, and 3)  groundtruth for the depth image. 
\\
\indent Using the SYNTHIA data set, we have simulated the common acquisition process of a mechanicaly scanned LiDAR system. We restrict the field of view (in elevation) of the  acquired depth, and add Gaussian noise with standard deviation that linearly increases with respect to the depth. Depth measurements  greater than a  maximum range were excluded, while training of parameters (using approximately 30 frames) for each of the evaluated methods were carried out using depth and intensity images separate from the test data. Finally, the random sampling levels used for obtaining  depth measurements were set as follows, 1) $6.25\%$  and 2) $1.56\%$.
\\
\indent The performance of our proposed method was evaluated aginst the following techniques, 1)  weighted TGV regularisation  \cite{Ferstl13}, 2)  the SD Filter \cite{Ham18} and 3) nearest neighbour interpolation. We assess the performance of the respective methods using  the following  quantitative metrics, the mean absolute error (MAE) and the root mean squared error (RMSE). Of the five SYNTHIA sequences, we used the following  for testing: sequence 1 spring (\textit{Highway}), sequence 2 fall (\textit{New York Like City}) and sequence 4 (\textit{Old European City}). The other two sequences were used for model parameter selection. The parameters used for all the SYNTHIA test sequences are  as follows: $M_{p}=5$, $\beta=0.005$ and $\gamma=0.001$. The number of iterations for each algorithm is varied such that the extracted depth is as close as possible to the true solution. The simulations of the proposed method were performed on a laptop with an Nvidia GeForce GTX 1060 graphics card, and our method took approximately 23 seconds per frame. 
\begin{figure*}[!t]
\captionsetup[subfigure]{}
\centering
\subfloat[Groundtruth]{\includegraphics[width=0.6\columnwidth]{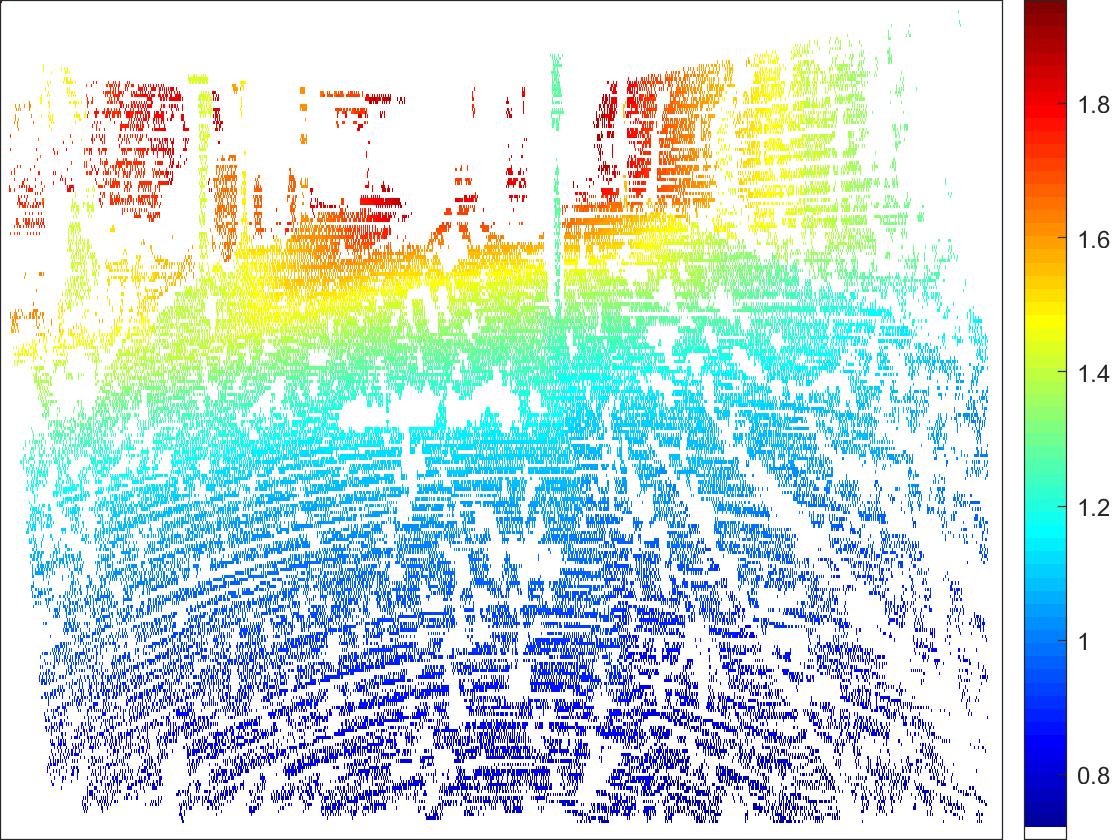}\label{fig:KITTI_GT}}
\quad
\subfloat[RGB Image]{\includegraphics[width=0.6\columnwidth]{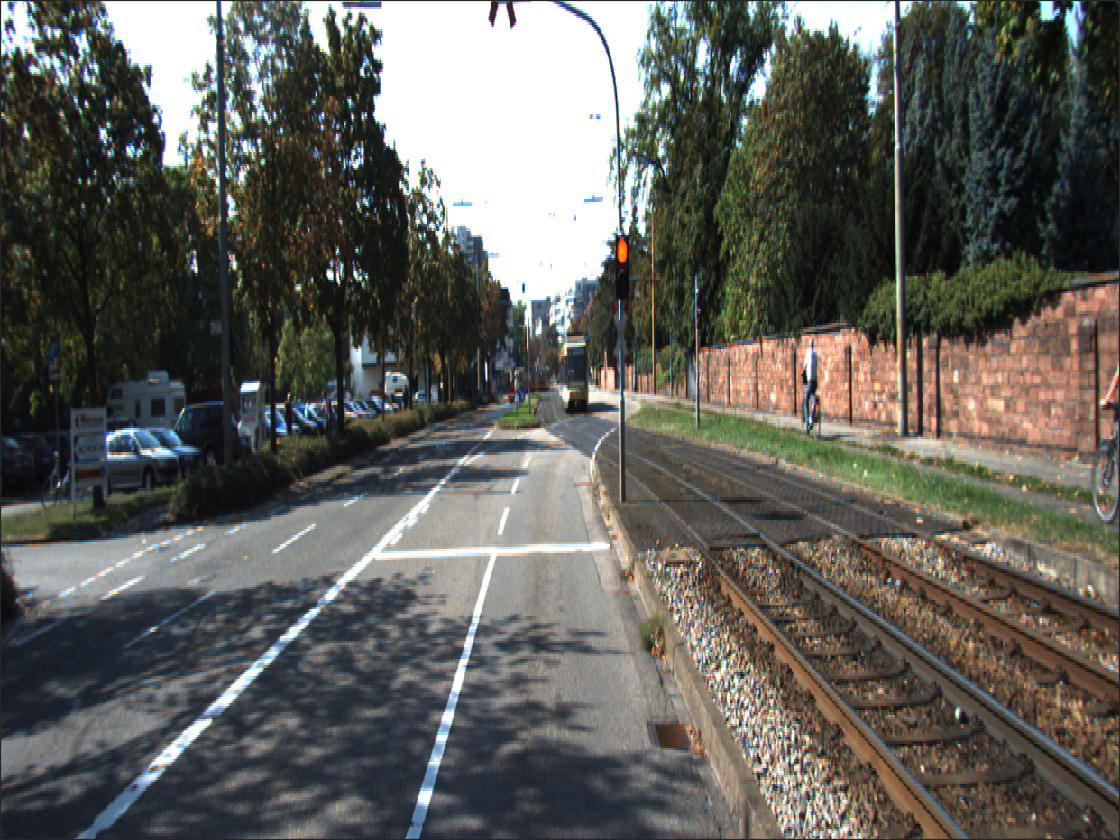}\label{fig:KITTI_Int}}
\quad
\subfloat[TGV \cite{Ferstl13}]{\includegraphics[width=0.6\columnwidth]{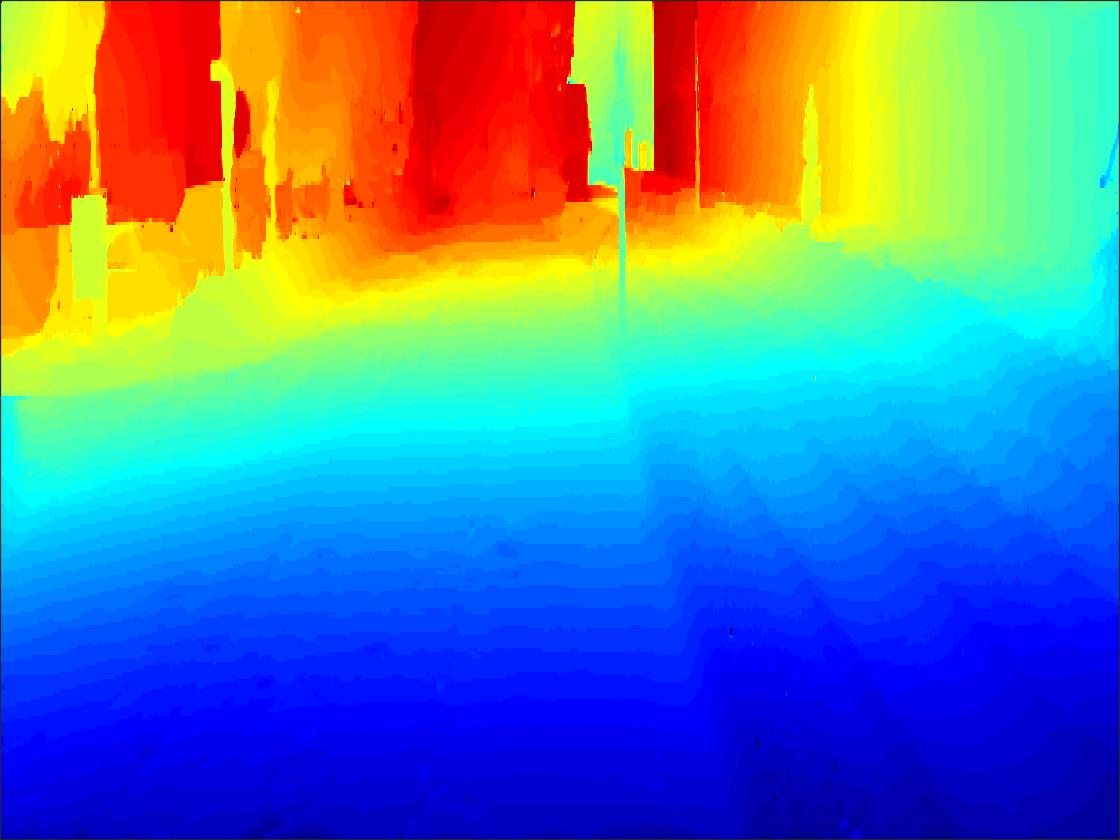}\label{fig:SynthiaRes_TGV}}
\\
\vspace{-2mm}
\subfloat[SD Filter \cite{Ham18} ]{\includegraphics[width=0.6\columnwidth]{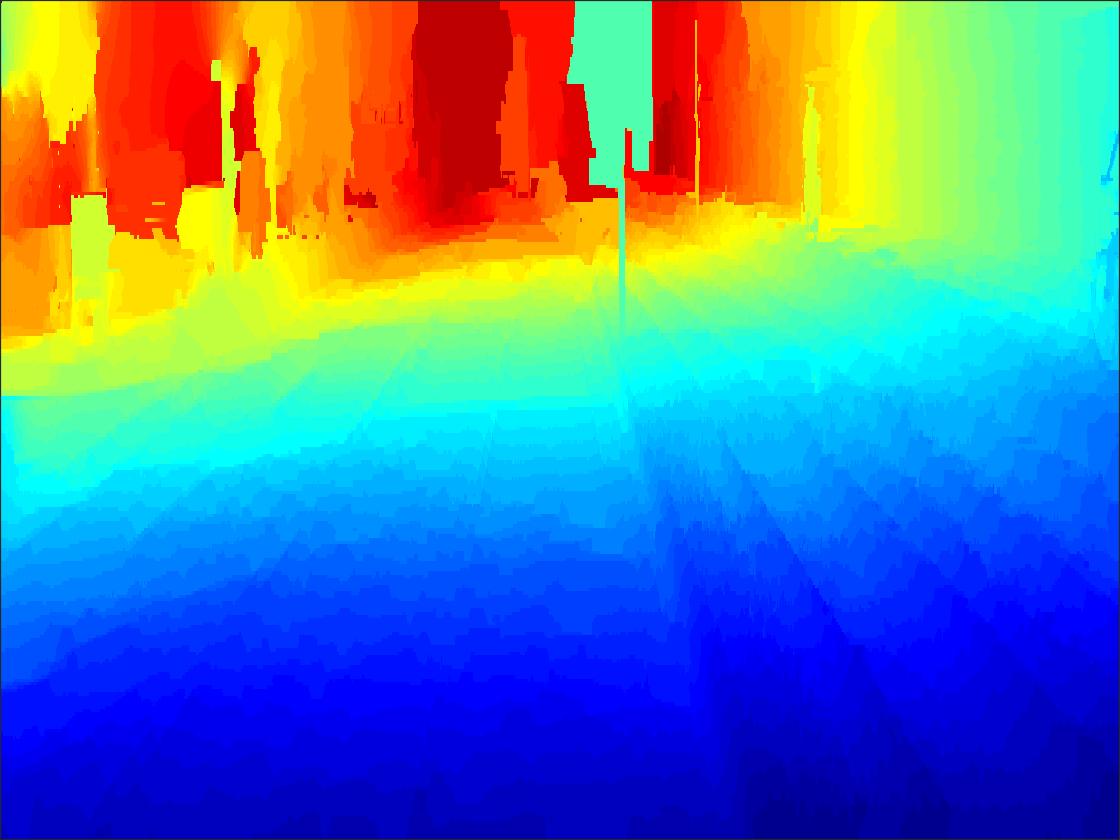}\label{fig:KITTI_SD}}
\quad
\subfloat[Proposed Method]{\includegraphics[width=0.6\columnwidth]{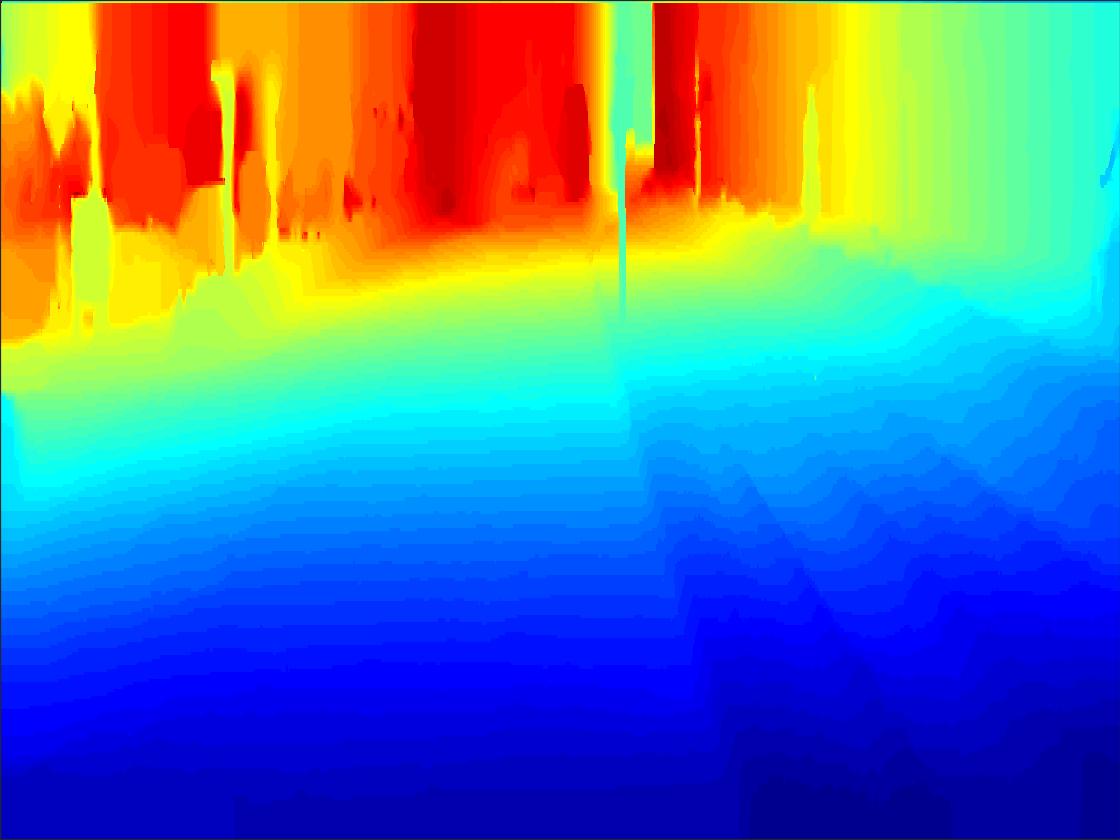}\label{fig:KITTI_PM}}
\quad
\subfloat[Baseline (without informing prior term)]{\includegraphics[width=0.6\columnwidth]{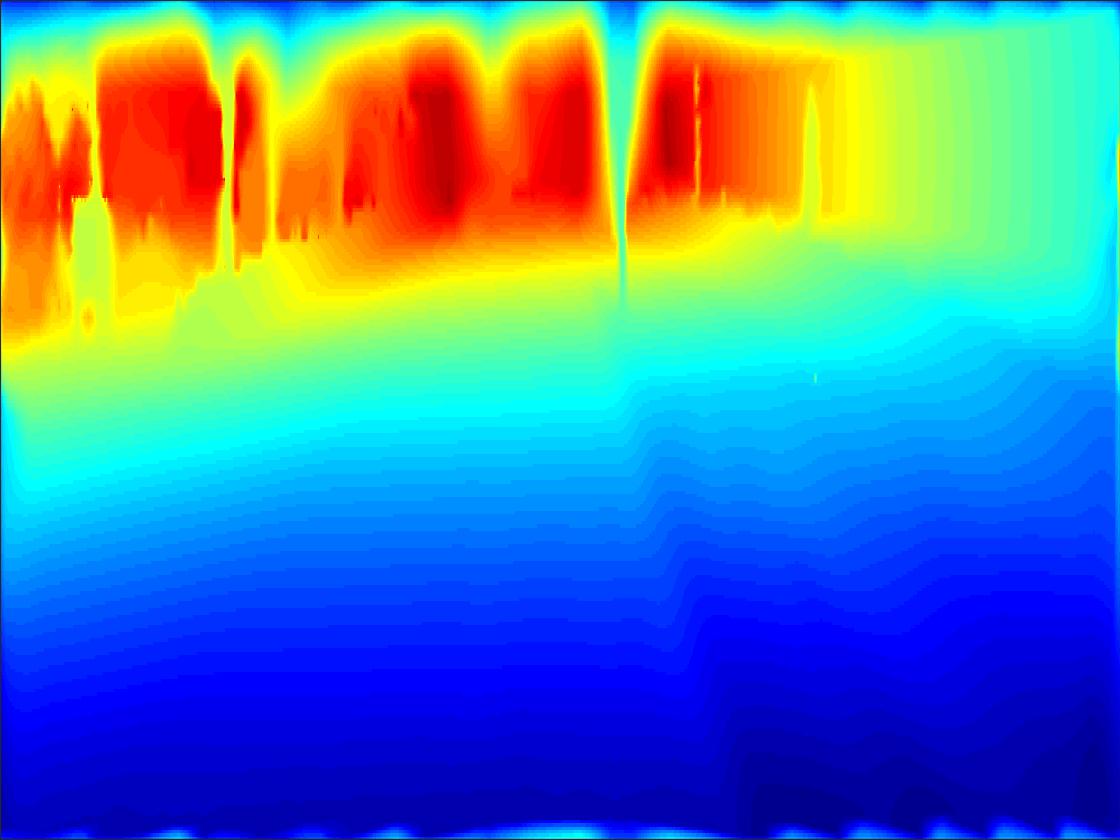}\label{fig:KITTI_Base}}
\caption{A qualitative comparison of a single instance of the KITTI dataset.  (a) The groundtruth depth. (b) The RGB image. (c) TGV \cite{Ferstl13}. (d) SD filter \cite{Ham18}. (e) Proposed method.  (f) Baseline depth upsampling. The logarithm of the depth values were used for colouration.}
\label{fig:KITTIres}
\end{figure*}
\\
\indent Table \ref{tab:Table1} shows the results of the respective methods, where a  performance improvement is defined  as a method having both the lowest MAE and RMSE errors. The proposed method was able to outperform the comparative methods in the \textit{Old European City} sequence (sparsity level of  $6.25\%$) as well as the \textit{New York Like City} sequence (at both  $6.25\%$ and $1.56\%$ sparsity levels). This implies that the use of the informing prior is able to enhance the recovery of the depth image, that is, the estimated magnitude and direction (sign) of a  change in depth aids  the upsampling process. However, the proposed method under-performed on the \textit{Highway} sequence at both the $6.25\%$ and $1.56\%$ sparsity levels, as well as at a sparsity of $1.56\%$ on the \textit{Old European City} sequence. This result  arises due to the following limitations with the proposed method: 1)  the number of edges being detected in the scene is low (as is the case in the \textit{Highway} sequence) and 2) there is a degradation in the estimated change in depth when the number of measurements obtained from the scene is low. 
\\
\indent A visual comparison of the respective methods is shown  in Fig.~\ref{fig:SynthiaRes}. The proposed method (as shown in Fig.~\ref{fig:SynthiaRes_PM}) is able to effectively remove noise while preserving depth discontinuities (especially when compared with the baseline shown in Fig.~\ref{fig:SynthiaRes_Base}). Furthermore, our method is more resilient  to texture variations, where this artifact can be observed in the estimated depth of the left stop sign (RGB image shown in Fig.~\ref{fig:SynthiaRes_Int}) for the TGV method (shown in Fig.~\ref{fig:SynthiaRes_TGV}) as well as to a lesser extent in the SD filter method (shown in Fig.~\ref{fig:SynthiaRes_SD}).
\subsection{KITTI Dataset}
We now evaluate the performance of the proposed method on the benchmarking KITTI  dataset \cite{Geiger2013} used to evaluate visual systems in relation to autonomous vehicles. The dataset used for depth upsampling consists of the following sensors, a  Velodyne-64E laser scanning system and a  Point Grey Flea 2 RGB camera. 
Sparse ground truth depth images (approximately 30\% complete) are provided to both train and evaluate the performance of the respective method; the ground truth is generated by the accumulation of the laser measurements  over a number of frames \cite{Uhrig17}.
\begin{table}[!b]
\centering
\normalsize
\begin{tabular}{llr}
\hline
    & MAE (m) & RMSE (m)\\
\hline
Proposed Method      & \textbf{0.43}    & \textbf{1.52}      \\
TGV \cite{Ferstl13}       & 0.43     & 1.67      \\
SD Filter \cite{Ham18}       & 0.46     & 1.91      \\
Nearest Neighbour     & 0.43    &1.86   \\    
\hline
\end{tabular}
\caption{Evaluation of the proposed method using the KITTI dataset.  The error (in metres) is measured by the MAE and RMSE. The best result is being highlighted}
\label{tab:Table2}
\end{table}
\\
\indent The depth  measurements are generated by projecting  the Velodyne-64E laser scans onto the  camera plane. We cropped the depth measurements along with the intensity images (carried out due to the  narrow field in elevation of the Velodyne-64E). Approximately 30 frames were used to determine the hyperparameters of the relevant models, while the evaluation was carried out using 7 sequences with approximately 850 frames. The average processing time of the proposed method was approximately 19 seconds per frame. The parameters are  as follows: $M_{p}=5$, $\beta=0.01$ and $\gamma=0.002$.
\\
\indent The averaged MAE and RMSE of the proposed method are shown in Table \ref{tab:Table2}. Note that MAE of the respective methods is approximately equal, with the proposed method having the lowest MAE error. However, the RMSE of the proposed method is significantly lower than the comparative methods, implying that the proposed method is able to preserve edges while recovering smooth surfaces for planar structures (such as the  ground plane). This can be observed in Fig.~\ref{fig:KITTIres}, where the proposed method is able to recover (similar to the TGV method \cite{Ferstl13}) a depth image with sharp edges; while also recovering smooth planar surfaces (e.g. the road) that are more resistant to depth image variations induced by intensity discontinuities that do not correspond to depth discontinuities. 
\section{Conclusions}
We have proposed an image guided depth upsampling method using   $\ell_{1}-\ell_{1}$ minimisation evaluated on both the SYNTHIA and KITTI datasets. Our method employed a linear combination of regularisers that also included a heuristic transformation  of the guiding image, referred to as the informing prior. A key advantage is that this informing prior is more robust to intensity discontinuities  that are not discontinuities in depth.   Furthermore, owing to our formulation we do not need large numbers of data frames for training. Future work will focus on improving the extraction of  discontinuities from the intensity image, as well as extending the optimisation model to incorporate data collected from radar. Such an extension would enable the development of a  robust depth upsampling method that has the potential to overcome performance degradation for optical sensors due to adverse environmental conditions. 
\section{Acknowledgement}
This work was supported by  Jaguar Land Rover and the UK-EPSRC grant EP/N012402/1 as part of the jointly funded Towards Autonomy: Smart and Connected Control (TASCC) Programme.
\bibliographystyle{IEEEtran}
\bibliography{ref}
\end{document}